\documentstyle[prl,aps,preprint]{revtex}

\begin{document}
\title{Searching for $S$-duality in  Gravitation\thanks{Invited talk at the {\it Third Workshop
on Gravitation and Mathematical-Physics,} Nov. 28-Dec. 3 1999, Le\'on Gto. M\'exico.}}
\author{H. Garc\'{\i}a-Compe\'an$^{a}$\thanks{%
E-mail: compean@fis.cinvestav.mx}, O. Obreg\'on$^{b}$\thanks{%
E-mail: octavio@ifug3.ugto.mx} and C. Ram\'{\i}rez$^c$\thanks{%
E-mail: cramirez@fcfm.buap.mx}}
\address{$^{a}$ {\it Departamento de F\'{\i}sica, Centro de Investigaci\'on y de}\\
Estudios Avanzados del IPN\\
P.O. Box 14-740, 07000, M\'exico D.F., M\'exico\\
$^b$ {\it Instituto de F\'{\i}sica de la Universidad de Guanajuato}\\
P.O. Box E-143, 37150, Le\'on Gto., M\'exico\\
$^c$ {\it Facultad de Ciencias F\'{\i}sico Matem\'aticas, Universidad}\\
Aut\'onoma de Puebla\\
P.O. Box 1364, 72000, Puebla, M\'exico}
\date{\today}
\maketitle

\begin{abstract}
\vskip-1.4truecm 

We overview some attempts to find $S$-duality analogues of non-supersymmetric
Yang-Mills theory, in the context of  gravity theories. 
The case of MacDowell-Mansouri gauge theory
of gravity is discussed. Three-dimensional dimensional reductions from the topological gravitational sector
in four dimensions, enable to recuperate the $2+1$ Chern-Simons gravity and the corresponding $S$-dual theory,
from the notion of self-duality in the four-dimensional theory.

\end{abstract}

\vskip -.5truecm





\newpage

\section{Introduction}

\setcounter{equation}{0}

Strong/weak coupling duality ($S$-duality) in superstring and supersymmetric gauge theories in
various dimensions has been, in the last five years, the major tool to study
the strong coupling dynamics of these theories. Much of these results require
supersymmetry through  the notion of BPS state. These states describe the physical spectrum 
and they are protected of 
quantum corrections leaving the strong/coupling duality under control to extract
physical information. In the non-supersymmetric case there are no BPS states and 
the situation is much more involved. This latter case is an open question and it is
still 
under current investigation.

In the specific case of non-supersymmetric gauge theories in four dimensions, the
subject has been explored recently in the Abelian as well as in the
non-Abelian cases \cite{witten,oganor} (for a review see \cite{quevedo}). In the
Abelian case, one considers $CP$ non-conserving Maxwell theory on a
curved compact four-manifold $X$ with Euclidean signature or, in other  
words, U(1) gauge theory with a $\theta$ vacuum coupled to
four-dimensional
gravity. The manifold $X$ is basically described by its associated
classical topological invariants: the Euler characteristic $\chi(X)  = { 1
\over 16 \pi^2}\int_X {\rm tr} R \wedge  \tilde R$ and the signature
$\sigma(X) = - {1 \over 24 \pi^2}\int_X {\rm tr} R\wedge R$.
In the Maxwell theory, the partition function $Z(\tau)$ transforms as a  
modular form under a finite index subgroup $\Gamma_0(2)$
of SL$(2,{\bf Z})$ \cite{witten}, $Z(-1/\tau) =
\tau^u \bar{\tau}^v Z(\tau)$, with the modular weight $(u,v) =
({1 \over 4}(\chi + \sigma), {1\over 4}(\chi - \sigma))$. In the above
formula $ \tau = {\theta \over 2 \pi} + {4 \pi i \over g^2}$,  where $g$   
is the U(1) electromagnetic coupling constant and $\theta$ is the usual
theta angle.

In order to  
cancel the modular anomaly in Abelian theories, it is known that one has to choose certain
holomorphic couplings $B(\tau)$ and $C(\tau)$ in  the
topological gravitational (non-dynamical) sector, through the action

\begin{equation}
 I^{TOP} = \int_X \bigg( B(\tau) {\rm tr} R \wedge \tilde R + C(\tau)
{\rm tr} R \wedge R \bigg),
\label{1}
\end{equation}
i.e., which is proportional to the appropriate sum of the Euler
characteristic
$\chi(X)$ and the signature $\sigma(X)$.

\vskip 1truecm 

\section{$S$-Duality in MacDowell-Mansouri Gauge Theory of Gravity}

Let us briefly review the MacDowell-Mansouri (MM) proposal \cite{mm}.  The starting point for the 
construction of
this theory is to consider an SO(3,2) gauge theory with a Lie
algebra-valued
gauge potential $A^{AB}_\mu$, where the indices $\mu = 0, 1, 2, 3$ are
space-time indices and the indices $A, B= 0, 1, 2, 3, 4$. From the gauge potential $A^{~AB}_\mu$ 
we may introduce the corresponding
field strength $F_{\mu\nu}^{~~AB} = \partial_\mu A_\nu^{~AB} - \partial_\nu A^{AB}_\mu
+ \frac{1}{2} f^{AB}_{CDEF} A^{CD}_\mu A^{EF}_\nu,$ 
where $f^{AB}_{CDEF}$ are the structure constants of SO(3,2).  MM
choose $F^{a4}_{\mu\nu} \equiv 0$ and as an action

\begin{equation}
S_{MM} = \int d^4 x \epsilon^{\mu\nu\alpha\beta} \epsilon_{abcd}
F^{~~ab}_{\mu\nu} F^{~~cd}_{\alpha\beta},
\end{equation}
where $a,b,...{\rm etc.}=0,1,2,3.$

On the other hand, by considering the self-dual (or
anti-self-dual) part of the connection, a generalization has been proposed
\cite{nos}. The extension to the supergravity case is considered in \cite{prl}.

One can then search whether the construction of a linear combination of
the corresponding self-dual and anti-self-dual parts of the
MacDowell-Mansouri action can be  reduced to the standard MM action
plus a kind of $\Theta$-term and, moreover, if by this means one can find
the ``dual-theory" associated with the MM theory.  This was showed in 
\cite{pursuing} and the corresponding extension to supergravity is given at \cite{dual}.
In what follows we follow Ref. \cite{pursuing}. Let us consider the action

\begin{equation}
S=\int d^4 x \epsilon^{\mu\nu\alpha\beta} \epsilon_{abcd} \bigg(
{^+} \tau {^+}F^{~~ab}_{\mu\nu} {^+}F^{~~cd}_{\alpha\beta} - {^-} \tau
{^-}F^{~~ab}_{\mu\nu} {^-}F^{~~cd}_{\alpha\beta} \bigg) ,
\end{equation}
where ${^\pm}F^{~~ab}_{\mu\nu} =  \frac{1}{2} \bigg( F^{~~ab}_{\mu\nu}
\pm \tilde F^{~~ad}_{\mu\nu} \bigg)$ and $\tilde F^{~~ab}_{\mu\nu} = - \frac{1}{2} i
\epsilon^{ab}_{~~cd}
F^{~~cd}_{\mu\nu}$.
It can be easily shown \cite{pursuing},
that this action can be rewritten as

\begin{equation}
S=\frac{1}{2} \int d^{4} x \epsilon^{\mu\nu\alpha\beta} \epsilon_{abcd}
\bigg[ ({^+}\tau - {^-}\tau) F^{~~ab}_{\mu\nu} F^{~~cd}_{\alpha\beta} + (
{^+}\tau + {^-}\tau) F^{~~ab}_{\mu\nu} \tilde F^{~~cd}_{\alpha\beta}
\bigg].
\end{equation}
In their original paper, MM have shown \cite{mm} that the first term in
this
action reduces to the Euler topological term plus the Einstein-Hilbert
action with a cosmological  term.  This was achieved after identifying
the components of the gauge field $A^{~AB}_{\mu}$ with the Ricci rotation
coefficients and the vierbein.   Similarly, the second term can be
shown  to be equal to $i\theta P$, where $P$ is the Pontrjagin topological
term \cite{nos}.  Thus, it is a genuine $\theta$ term,  with $\theta$
given by the sum ${^+}\tau + {^-}\tau$.

Our second task is to find the ``dual theory", following the same scheme
as for
Yang-Mills theories \cite{oganor}.  For that purpose
we consider the parent action

\begin{equation}
I= \int d^{4}x \epsilon^{\mu\nu\alpha\beta} \epsilon_{abcd}
\bigg( c_1{^+} G^{~~ab}_{\mu\nu} {^+}G^{~~cd}_{\alpha\beta} + c_2
{^-} G^{~~ab}_{\mu\nu} {^-} G^{~~cd}_{\alpha\beta} + c_3 {^+}
F^{~~ab}_{\mu\nu} {^+}G^{~~cd}_{\alpha\beta} + c_4 {^-} F^{~~ab}_{\mu\nu}
{^-} G^{~~cd}_{\alpha\beta} \bigg) .
\end{equation}
From which the action (3) can be
recovered after integration on ${^+}G$ and ${^-}G$.

In order to get the ``dual theory" one should start with the partition function

\begin{equation}
Z= \int {\cal D} {^+}G \, {\cal D} {^-}G \, {\cal D} A\,\,e^{-I}.
\end{equation}
To proceed with  the integration over the gauge fields we observe that
$F^{~~ab}_{\mu\nu} = \partial_\mu A^{~ab}_{\nu} - \partial_\nu A^{~ab}_\mu
+ \frac{1}{2} f^{ab}_{CDEF} A^{~CD}_{\mu} A^{~EF}_{\nu}.$ 
Taking into account the explicit
expression for the structure constants, the second term of $F^{~~ab}_{\mu\nu}$ 
will naturally split in
four terms given by $A^{~ad}_{\mu} A^{~~b}
_{\nu d}$ $- A^{~ad}_\nu A^{~~b}_{\mu d}$ $ - \lambda^2 \bigg( A^{~a4}_\mu
A^{~b4}_\nu - A^{~a4}_\nu A^{~b4}_\mu \bigg).$ The integration over the  
components $A^{~a4}_\mu$ is given by a  Gaussian integral, which turns out
to be $det\, {\bf G}^{-1/2}$, where  {\bf G} is a matrix given by
${\bf G}^{~~\mu\nu}_{ab}$ $ = 8 i \lambda^2 \epsilon^{\mu\nu\alpha\beta}
\bigg( c_3 {^+}G_{\alpha\beta ab} - c_4 {^-}G_{\alpha\beta ab} \bigg).$

Thus, the partition function (6) can be written as

\begin{equation}
Z= \int {\cal D}{^+} G\, {\cal D} {^-}G \, {\cal D} A^{~ab}_\mu\,\, det\,
{\bf
G}^{-1/2}\,\, e^{- \, {I\!\!I}},
\end{equation}
where

$${I\!\!I}=2 i \int d^4 x  \epsilon^{\mu\nu\alpha\beta} \bigg[ c_1
{^+}G^{~~ab}_{\mu\nu} {^+}G_{\alpha\beta ab} - c_2 {^-} G^{~~ab}_{\mu\nu}
{^-}G_{\alpha\beta ab} + 2 H^{~~ab}_{\mu\nu} ( c_3
{^+} G_{\alpha\beta ab} - c_4 {^-}G_{\alpha\beta ab}) \bigg], $$
and $H^{~~ab}_{\mu\nu} = \partial_\mu A^{~ab}_\nu - \partial_\nu A^{~ab}_\mu
+ \frac{1}{2} f^{ab}_{cdef} A^{~cd}_\mu A^{~ef}_\nu$
is the SO(3,1) field strength.

Our last step to get the dual action is to integrate over
$A^{~ab}_\mu$.  This kind of integration is well known and has been   
performed in previous works \cite{oganor,pursuing,towards}.  The result is

\begin{equation}
Z=\int {\cal D}{^+}G\, {\cal D} {^-}G \,\, det\, {\bf G}^{-1/2} \,\,
det ({^+}M)^{-1/2} det ({^-}M)^{-1/2} \,\, e^{- \int d^4x \tilde L},
\end{equation}
with

\begin{equation}
\begin{array}{ll}
\tilde{L} &= \epsilon^{\mu\nu\rho\sigma} \bigg[ -{1 \over 4
{^+}\tau} {^+}G^{~~ab}_{\mu \nu} {^+}G_{\rho\sigma ab} + {1 \over 4
{^-}\tau} {^-}G^{~~ab}_{\mu \nu} {^-}G_{\rho \sigma ab} +
2 \partial_{\nu} {^+}G_{\rho \sigma ab} {({^+}M)}^{-1 \ abcd}_{\mu
\lambda} \epsilon^{\lambda \theta \alpha \beta}
\partial_{\theta} {^+}G_{\alpha \beta cd} \\
&- 2 \partial_{\nu} {^-}G_{\rho
\sigma
ab} {({^-}M)}^{-1 \ abcd}_{\mu
\lambda} \epsilon^{\lambda \theta \alpha \beta}
\partial_{\theta} G^-_{\alpha \beta cd}\bigg],
\end{array}
\label{tcinco}
\end{equation}
where ${^\pm}M^{\mu\nu ~cd}_{~~ab} = \frac{1}{2} \epsilon^{\mu\nu\alpha\beta}
\bigg( - \delta^c_a {^\pm}G^{~~~d}_{\alpha\beta b} + \delta^c_b
{^\pm}G^{~~~ d}_{\alpha\beta a} + \delta^d_a {^\pm} G^{~~~ c}_{\alpha\beta
b}
- \delta^d_b {^\pm} G^{~~~ c}_{\alpha\beta a} \bigg)$ and ${^+} \tau = - \frac{1}{4c_1},$ \
${^-}\tau = - \frac{1}{4c_2}$,
$c_3 = c_4 = 1.$

The non-dynamical model considered in a previous work \cite{towards} 
results in a kind of non-linear sigma model \cite{oganor} of the type 
considered by Freedman and
Townsend \cite{freed}, 
as in the usual Yang-Mills dual models.  The
dual to the dynamical gravitational model (9) considered here, 
results in a Lagrangian of the same
structure.  However, it differs from the non-dynamical case by the 
features discussed above.

\vskip 1truecm 

\section{(Anti)Self-duality of the Three-dimensional Chern-Simons Gravity }

It is well known that the $2+1$ Einstein-Hilbert action
with nonvanishing cosmological constant $\lambda$ is given by the 
``standard'' and ``exotic'' Einstein actions \cite{wittenone}. It is well known that for 
$\lambda >0$ (and $\lambda <0$),
these actions are equivalent to a Chern-Simons actions
in $2+1$ dimensions with gauge group ${\cal G}$ to be SO(3,1) (and SO(2,2)). 

In this section we will work out the Chern-Simons Lagrangian for (anti)self-dual
gauge connection with respect to duality transformations of the internal
indices of the gauge group ${\cal G}$, in the same philosophy of MM \cite{mm},
and that of \cite{nos}

\begin{equation}
L{^{\pm}}_{CS} = \int_{\cal M} \varepsilon^{ijk} \bigg( {^{\pm}} A_i^{AB}
\partial_j {\ ^{\pm}} A_{kAB} + {\frac{2 }{3}} {^{\pm}} A_{iA}^B {^{\pm}}
A_{jB}^C {^{\pm}} A_{kC}^A \bigg),  \label{csself}
\end{equation}
where $A,B,C,D= 0,1,2,3,$ $\eta_{AB} = diag(-1,+1,+1,+1)$ and the complex (anti) 
self-dual connections are ${^{\pm}} A_{i}^{AB} = {\frac{1}{2}}(A_{i}^{AB} \mp {\frac{i}{2}}
\varepsilon^{AB}_{ \ \ CD} A_{i}^{CD})$, which satisfy the relation $\varepsilon^{AB}_{ \ \ CD}
{^{\pm}} A_i^{CD} = \pm i {^{\pm}} A^{AB}.$

Thus using the above equations, the action (10) can
be rewritten as

\begin{equation}
L{^{\pm }}_{CS}=\int_{\cal M}{\frac{1}{2}}\varepsilon ^{ijk}\bigg(
A_{i}^{AB}\partial _{j}A_{kAB}+{\frac{2}{3}}A_{iA}^{\ \ B}A_{jB}^{\ \
C}A_{kC}^{\ \ A}\bigg)\mp {\frac{i}{4}}\varepsilon ^{ijk}\varepsilon ^{ABCD}
\bigg(A_{iAB}\partial _{j}A_{kCD}+{\frac{2}{3}}A_{i\text{ }A}^{\
E}A_{jEB}A_{kCD}\bigg).  \label{eq13}
\end{equation}

\bigskip

In this expression the first term is the Chern-Simons action for the gauge
group ${\cal G}$, while the second term appears as its corresponding ``$
\theta $-term''. The same result was obtained in $3+1$ dimensions when we
considered the (anti)self-dual MM action \cite{pursuing}, or the
(anti)self-dual $3+1$ pure topological gravitational action \cite{towards}.  

One should remark that the two terms in the action (13)  are the
Chern-Simons and the corresponding ``$\theta $-term'' for the gauge group
${\cal G}$ under consideration. After imposing the particular
identification 
$A_i^{AB}=(A_{i}^{ab},A_{i}^{3a})=(\omega_{i}^{ab},\sqrt{\lambda
}e_{i}^{a})$ and $\omega_{i}^{ab}=\varepsilon^{abc} \omega_{ic}$, the
``exotic'' and ``standard'' actions for the gauge group SO(3,1) are
given respectively by

\[
L_{CS}{^{\pm}} = \int_X {\frac{1}{2}} \varepsilon^{ijk} \bigg(
\omega^a_i(\partial_j \omega_{ka} - \partial_k \omega_{ja}) + {\frac{2 }{3}}
\varepsilon_{abc} \omega_i^a \omega_j^b \omega_k^c + \lambda
e^a_i(\partial_j e_{ka} - \partial_k e_{ia}) - 2 \lambda
\varepsilon_{abc}e_i^a e_j^b \omega_k^c\bigg)
\]

\begin{equation}
\pm i \sqrt{\lambda}\varepsilon ^{ijk}\bigg(e_{i}^{a}(\partial _{j}\omega _{ka}-\partial
_{k}\omega _{ja})-\varepsilon _{abc}e_{i}^{a}\omega _{j}^{b}\omega _{k}^{c}+{
\frac{1}{3}}\lambda \varepsilon _{abc}e_{i}^{a}e_{j}^{b}e_{k}^{c}\bigg),
\label{eq14}
\end{equation}
plus surface terms. It is interesting to note that the above action (12) can be obtained
from action (1) (for a suitable choice of $B(\tau)$ and $C(\tau)$) by 
dimensional reduction from $X$ to its boundary ${\cal M} = \partial X$. Thus the ``standard'' 
action come from the Euler characteristic $\chi(X)$, while the ``exotic'' action come from 
the signature $\sigma(X).$

\vskip 1truecm 

\section{Chern-Simons Gravity Dual Action in Three Dimensions}

This section is devoted to show that a ``dual'' action to the Chern-Simons
gravity action can be constructed following \cite{sabido}. Essentially we will
repeat the procedure to find the the ``dual'' action to MM gauge theory
given in Sec. II.

We begin from the original non-Abelian Chern-Simons action given by

\begin{equation}
L=\int_{\cal M}d^{3}x{\frac{g}{4\pi }}\varepsilon ^{ijk}A_{i}^{AB}\bigg(\partial
_{j}A_{kAB}+{\frac{1}{3}}f_{ABCDEF}A_{j}^{CD}A_{k}^{EF}\bigg).  \label{cs}
\end{equation}
Now, as usual we propose a parent action in order to derive the dual action

\begin{equation}
L_D = \int_{\cal M} d^3 x \varepsilon^{ijk} \bigg( a B^{AB}_i H_{jkAB} + b A^{AB}_i
G_{jkAB} + c B^{AB}_i G_{jkAB} \bigg),  \label{parent}
\end{equation}
where $H_{jkAB}= \partial_j A_{kAB} + {\frac{1 }{3}} f_{ABCDEF}A^{CD}_jA^{EF}_k$ 
and $B^{AB}_i$ and $G^{AB}_{ij}$ are vector and tensor fields on ${\cal M}$.
It is a very easy matter to show that the action (13) can be derived from this
parent action after integration of $G$ fields

The ``dual'' action $L^*_D$ can be computed as usually in the Euclidean
partition function, by integrating first out with respect to the physical
degrees of freedom $A^{AB}_i.$ 
The resulting action is of the Gaussian type in the variable $A$ and thus,
after some computations, it is easy to find the ``dual'' action

\begin{equation}
L_{D}^{\ast }=\int_{\cal M}d^{3}x\varepsilon ^{ijk}\bigg \{-{\frac{3}{4a}}
(a\partial _{i}B_{jAB}+bG_{ijAB})[{\bf R}^{-1}]_{kn}^{ABCD}\varepsilon
^{lmn}(a\partial _{l}B_{mCD}+bG_{lmCD})+c\alpha _{i}^{AB}G_{jkAB}\bigg\},
\label{bosonicdual}
\end{equation}
where $[{\bf R}]$ is given by $[{\bf R}]_{ABCD}^{ij}=\varepsilon ^{ijk}f_{\
\ ABCD}^{EF}B_{kEF}$ whose inverse is defined by $[{\bf R}]_{ABCD}^{ij}[{\bf 
R}^{-1}]_{jk}^{CDEF}=\delta _{k}^{i}\delta _{AB}^{EF}.$

The partition function is finally given by 
\begin{equation}
Z = \int {\cal D}G {\cal D}B \sqrt{det({\bf M}^{-1})} exp \big( - L^*_D
\big).
\end{equation}
In this ``dual action'' the $G$ field is not dynamical and can be
integrated out. The integration of this auxiliary
field gives 
\begin{equation}
L^{**}_D=\int_{\cal M} d^3 x{\frac{4\pi }{g}}\varepsilon
^{lmn} \bigg(B_{l}^{AB}\partial _{m}B_{nAB}-{\frac{4\pi }{g}}
f_{ABCDEF}B_{l}^{AB}B_{m}^{CD}B_{n}^{EF}\bigg).
\end{equation}
The fields $B$ cannot be rescaled if we impose ``periodicity'' conditions 
on them. Thus, this
dual action has inverted coupling with respect to the original one (compare with
\cite{bala} for the Abelian case). 

\newpage

\vskip 2truecm
\centerline{\bf Acknowledgments}
The results of sections 3 and 4 were obtained in collaboration with Miguel Sabido.
We are very grateful to him for a critical reading of this manuscript.

\vskip 1truecm 



\begin{references}

\bibitem{witten}  E. Witten, Selecta Mathematica, New Series, {\bf 1},
383 (1995).

\bibitem{oganor}  O. Ganor and J. Sonnenschein, Int. J. Mod. Phys. A {\bf 11}
(1996) 5701; N. Mohammedi, hep-th/9507040; Y. Lozano, Phys. Lett. B {\bf 364}
(1995) 19.

\bibitem{quevedo}  F. Quevedo, Nucl. Phys. Proc. Suppl. {\bf 61} A (1998) 23.

\bibitem{mm}  S.W. Mac Dowell and F. Mansouri, Phys. Rev. Lett. {\bf 38}
(1977) 739; F. Mansouri, Phys. Rev. D {\bf 16} (1977) 2456.

\bibitem{nos}  J.A. Nieto, O. Obreg\'{o}n and J. Socorro, Phys. Rev. D {\bf %
50} (1994) R3583.

\bibitem{prl} J.A. Nieto, J. Socorro and O. Obreg\'on, Phys. Rev. Lett. {\bf 76}
(1996) 3482.  

\bibitem{pursuing}  H. Garc\'{\i}a-Compe\'{a}n, O. Obreg\'{o}n and C.
Ram\'{\i}rez, Phys. Rev. D {\bf 58}, 104012-1 (1998); Chaos,
Solitons and Fractals {\bf 10} (1999) 373.

\bibitem{dual} H. Garc\'{\i}a-Compe\'{a}n, J.A. Nieto, O. Obreg\'{o}n and C. Ram\'{\i}rez,
Phys. Rev. D {\bf 59} (1999) 124003.

\bibitem{towards}  H. Garc\'{\i}a-Compe\'{a}n, O. Obreg\'{o}n, J.F.
Pleba\'{n}ski and C. Ram\'{\i}rez, Phys. Rev. D {\bf 57}, 7501 (1998).

\bibitem{freed}  D. Z. Freedman and P. K. Townsend, Nucl. Phys. B{\bf \ 177}%
, 282 (1981); P. C. West, Phys. Lett. B {\bf 76}, 569 (1978); A. H.
Chamseddine, Ann. Phys. (N.Y.) {\bf 113}, 219 (1978); S. Gotzes and A. C.
Hirshfeld, {\it ibid}. {\bf 203}, 410 (1990).

\bibitem{wittenone}  E. Witten, Nucl. Phys. B {\bf 311} (1988) 46; 
Nucl. Phys. B {\bf 323} (1989) 113.

\bibitem{sabido}  H. Garc\'{\i}a-Compe\'{a}n, O. Obreg\'{o}n, C. Ram\'{\i}rez
and M. Sabido, ``Remarks on $2+1$ Self-dual Chern-Simons Gravity'', hep-th/9906154,
to appear in Phys. Rev. D. 

\bibitem{bala}  A.P. Balachandran, L. Chandar and B. Sathiapalan, Int. J. Mod. Phys. A 
{\bf 11} (1996).



\end{references}
\end{document}